\documentclass[journal=jctcce,manuscript=article]{achemso}

\usepackage[version=3]{mhchem} % Formula subscripts using \ce{}
\usepackage[dvipsnames]{xcolor}
\usepackage{graphicx,subcaption}
\usepackage{amssymb}
\usepackage{braket}
\usepackage{bm}
\usepackage{derivative}
\usepackage{hyperref}
\usepackage{siunitx}
\usepackage{booktabs}
\usepackage{bigstrut}
\usepackage{multirow}

\author{William Colglazier}
\affiliation{Theoretical Division, Los Alamos National Laboratory, Los Alamos, NM, USA}
\email{maxim@lanl.gov}

\author{Nicholas Lubbers}
\affiliation{Computer, Computational, and Statistical Sciences Division, Los Alamos National Laboratory, Los Alamos, NM, USA}

\author{Sergei Tretiak}
\affiliation{Theoretical Division, Los Alamos National Laboratory, Los Alamos, NM, USA}
\alsoaffiliation{Center for Integrated Nanotechnologies, Los Alamos National Laboratory, Los Alamos, NM, USA}
\alsoaffiliation{Center for Nonlinear Studies, Los Alamos National Laboratory, Los Alamos, NM, USA}

\author{\\Anders M. N. Niklasson}
\affiliation{Theoretical Division, Los Alamos National Laboratory, Los Alamos, NM, USA}

\author{Maksim Kulichenko}
\affiliation{Theoretical Division, Los Alamos National Laboratory, Los Alamos, NM, USA}
\email{wcolglazier@lanl.gov}

\title{Enhancing Molecular Dipole Moment Prediction with Multitask Machine Learning}
\begin{document}

% \begin{tocentry}

% Some journals require a graphical entry for the Table of Contents.
% This should be laid out ``print ready'' so that the sizing of the

% Inside the \texttt{tocentry} environment, the font used is Helvetica
% 8\,pt, as required by \emph{Journal of the American Chemical

% The surrounding frame is 9\,cm by 3.5\,cm, which is the maximum
% permitted for  \emph{Journal of the American Chemical Society}
% graphical table of content entries. The box will not resize if the
% content is too big: instead it will overflow the edge of the box.

% This box and the associated title will always be printed on a
% separate page at the end of the document.

% \end{tocentry}

\begin{abstract}
We present a multitask machine learning strategy for improving the prediction of molecular dipole moments by simultaneously training on quantum dipole magnitudes and inexpensive Mulliken atomic charges. With dipole magnitudes as the primary target and assuming only scalar dipole values are available without vector components we examine whether incorporating lower quality labels that do not quantitatively reproduce the target property can still enhance model accuracy. Mulliken charges were chosen intentionally as an auxiliary task, since they lack quantitative accuracy yet encode qualitative physical information about charge distribution. Our results show that including Mulliken charges with a small weight in the loss function yields up to a 30\% improvement in dipole prediction accuracy. This multitask approach enables the model to learn a more physically grounded representation of charge distributions, thereby improving both the accuracy and consistency of dipole magnitude predictions. These findings highlight that even auxiliary data of limited quantitative reliability can provide valuable qualitative physical insights, ultimately strengthening the predictive power of machine learning models for molecular properties.
\end{abstract}


 \section{Introduction}
Modeling molecular properties such as energies, forces, dipole moments, and other quantities with quantum level accuracy is of great importance in computational chemistry and physics, as these properties and their interrelations provide crucial insights into the processes occurring within and between molecules and materials.~\cite{szabo1996modern,jensen2017intro,schutt2019unifying,unke2019physnet,veit2020muml,perea2018dipoleml,zhang2022gnnchem,qiao2020orbnet,chen2022chargenet,li2024dipoleml,thaler2024activecharges,hernmann2022abinitiowavefn,unk2021spookynet}. When modeling molecular systems, a variety of methods are available, each with their own trade off between accuracy and computational cost~\cite{cramer2013essentials,mardirossian2017thirty,harris2021mlpsi,zhang2021forcecharge,amano2024mldipoleglycol,ma2024pretrainedE3GNN,qian2024unifiedGW,villani2024hamiltonianml,li2025shadowflexiblecharge}. For example, one of the most popular methods, Density Functional Theory (DFT), offers a relatively accurate and efficient approach, with a computational cost of $\mathcal{O}(n^3)$, where $n$ is proportional to the number basis functions in the system~\cite{10.1007/978-94-009-9027-2_2,RevModPhys.87.897,xu2024egeognn,smith2025hessianmlip,chen2023electronpropertiesGNN,yan2025QGEMcharge,wang2024deepnosearchgpu,muller2020MLforcefields,behler2021gapreview,ko2024chargequellayer}. A quantum chemistry golden standard Coupled Cluster with Singles, Doubles, and perturbative Triples (CCSD(T)) is significantly more computationally expensive, scaling at $\mathcal{O}(n^7)$~\cite{RevModPhys.79.291}. 
% In more practical applications, such as high throughput screening where the properties of millions of molecules are calculated to find the best ones for a specific use cases traditional quantum chemistry methods can often be too slow and computationally expensive for this
In practical applications like high throughput screening where researchers are aiming to identify promising molecular candidates for specific applications which often requires the evaluation of molecular properties for tens of millions of molecules making these quantum chemistry methods in a lot of use cases far too computationally expensive and time consuming to be applied at this scale.~\cite{pyzerknapp2015htvs,vonlilienfeld2013compound,kulichenko2023nature,kulichenko2021synergy,doe2025momentGNN,yuan2025QDFTnet,johnson2025chargeGNN,smith2022deepHam,athavale2025pyseqm20acceleratedsemiempirical}. In addition, supervising atomic charges provides a direct, chemically meaningful per-atom quantity that can be inspected, making the model more interpretable and less of a black box.

Recent advances in Machine Learning (ML) have demonstrated the ability to achieve accuracy comparable to traditional quantum chemistry methods, while dramatically lowering computational costs and achieving linear $\mathcal{O}(n)$ scaling~\cite{schutt2017schnet,xie2018crystal,kim2023schrgnn,park2023equivariantgnn,pasi2023mlpsi,park2023equivariantml,kim2023schnrGNN,li2024hybridGNNdipole}. However, while many ML studies have successfully predicted extensive properties such as total molecular energies, intensive properties such as HOMO–LUMO gaps or dipole moments pose greater challenges. Because these properties are size independent and governed by spatial nonlocality and long range interactions, standard ML models that rely on local atomic environments often struggle to capture them. This motivates the need for approaches that embed additional physical information learning signals. To accomplish this, various types of ML architectures can be employed, such as message passing neural networks (MPNNs)~\cite{gilmer2017neural}, graph neural networks (GNNs)~\cite{kipf2016semi,scarselli2008graph}, or transformer based models~\cite{wang2019reinforced}, all of which can be trained to predict quantum mechanically computed properties~\cite{noe2020machine,carleo2019machine,kulichenko2023rise,kulichenko2022extending}.

ML models that can predict such quantum chemical data with high accuracy and computational efficiency are of especial interest in fields of drug discovery~\cite{ragoza2017protein}, semiconductors~\cite{pilania2013accelerating}, and material design~\cite{huan2017universal}. However, one of key challenges of ML approaches is to achieve extensibility and transferability of a model while minimizing the inference error~\cite{kulichenko2023nature}. One approach to improve this is to use multitask learning, where a single model is trained to perform predictions for multiple related tasks. Previous studies have shown that multitask learning can enhance the performance of ML models by using shared representations of related properties. For instance, Gastegger et al.~\cite{gastegger2017machine} trained NN to simultaneously predict atomic forces and dipole vectors, demonstrating improved generalization. Schütt et al.~\cite{schutt2019quantum} developed PhysNet, a model that predicts energies, forces, and dipoles using hierarchical physical constraints. Other efforts~\cite{schwalbekoda2022generative} have incorporated auxiliary tasks such as charge or electron density prediction to support global molecular property estimation. However, most of these models treat atomic charges as latent intermediate representations rather than directly supervising them.

In this work, we introduce a multitask ML approach that jointly learns atomic charges and dipole magnitudes. That is, we simulate the scenario when only scalar dipole magnitudes are available without their vector components. Under this assumption, we investigate if the inclusion of data that lacks quantitative accuracy but provides important qualitative physics can improve the ML model's performance. Our method differs by explicitly training mainly on dipole magnitudes and treating atomic charge prediction as a secondary task. This encourages the model to develop more physically meaningful atomic representations that better reflect the underlying charge distribution in molecules~\cite{acs2023jctc,nature2023natm}. For the atomic charges, we use Mulliken charges, an important choice because they are one of the least computationally expensive charge partitioning schemes available in quantum chemistry~\cite{acs2021jcp}. Although Mulliken charges are known to be inaccurate, their inclusion provides valuable atomic level information at a negligible computational cost which helps guide the model toward learning a more physically consistent understanding of molecular polarity, ultimately improving dipole prediction accuracy and robustness~\cite{aip2023overview,acs2024review,nature2022review}. That is, we deliberately choose this type charges to highlight that, despite the quantitative accuracy of an auxiliary label, multitask learning improves outcomes

% \newpage

% In this work, we introduce a multitask machine learning approach that jointly learns atomic charges and scalar dipole magnitudes. In our method, we train primarily on dipole magnitudes and treat atomic charge prediction as an additional task. This helps encourage the model to learn more physically meaningful atomic representations that capture the true charge distribution~\cite{acs2023jctc,nature2023natm}. Importantly, we use the most computationally affordable charges available in QM calculations—Mulliken charges—which by themselves do not properly reproduce quantum dipoles~\cite{acs2021jcp}. However, in joint training with quantum dipole data, they play a key role in improving the accuracy and physicality of the model. We find that including charge prediction with a small loss weight of just 4\% leads to over a 30\% improvement in dipole magnitude accuracy. This result highlights the value of simple, physically interpretable auxiliary labels~\cite{aip2023overview} and demonstrates how incorporating atomic-level information can enhance the prediction of global molecular properties in a data-efficient and physically grounded way~\cite{acs2024review,nature2022review}.

% Our approach differs by explicitly incorporating partial charges as an auxiliary task alongside dipole magnitudes in the loss function while the dipole is explicitly defined as a function of predicted charges. 

\section{Methods}
% \subsectionx{Theory}

Dipoles are a per molecule global feature, which can either be represented as an XYZ vector or a scalar magnitude that measures how the electric charge is distributed between atoms across a molecule, resulting in one region having a partial positive charge and another a partial negative charge. This phenomenon typically arises upon the assmbly of atoms with differing electronegativities, leading to an uneven distribution of electron density~\cite{tuckerman2006quantum}.The resulting charge asymmetry is quantified by the molecular dipole moment. While the dipole captures the overall polarity of a molecule, atomic charges provide a localized view of the electron distribution at individual atoms. Among the many charge partitioning schemes in quantum chemistry we choose to work withthe Mulliken population analysis which computes the atomic charge \( q_A \)~\cite{mulliken1955electronic} as:

\begin{equation}
q_A = Z_A - \sum_{\mu \in A} P_{\mu\mu} - \sum_{\mu \in A} \sum_{\nu \notin A} P_{\mu\nu} S_{\mu\nu}
\label{eq:partial_mulliken}
\end{equation}

Here, \( Z_A \) is the nuclear charge of atom \( A \), \( P_{\mu\nu} \) is the density matrix element between atomic orbitals \( \mu \) and \( \nu \), and \( S_{\mu\nu} \) is the corresponding overlap matrix element. The indices \( \mu \in A \) refer to basis functions centered on atom \( A \), and \( \nu \notin A \) refer to basis functions centered on atoms other than \( A \). The first summation term, \( \sum_{\mu \in A} P_{\mu\mu} \), captures the electron density localized on atom \( A \), while the second double summation accounts for the shared electron density between atom \( A \) and its neighboring atoms via orbital overlap. By subtracting both from the nuclear charge, the Mulliken analysis provides an estimate of the net atomic charge \( q_A \). There are many forms of charge partitioning schemes that can be applied to calculate atomic charges for molecular systems, all with their own trade off of computational cost vs accuracy. Among some of the most popular charge partitioning schemes including Mulliken, Hirshfeld, and NBO. In this work, we choose to use Mulliken charges because they are among the least computationally expensive methods, though they are also known to be among the least accurate for analysis of charge distribution across molecular structure. This inaccuracy becomes particularly evident when we calculate the dipole moment \( \mu \) using the point charge approximation\citep{Hirshfeld1977,Reed1985,Sigfridsson1998}

% Among many schemes (Mulliken, Hirshfeld/stockholder, and NBO/NPA, among others), we chose to work with Mulliken partial charges because they are one of the least computationally expensive methods, though they are also known to be the one of the least accurate. This inaccuracy becomes particularly evident when we calculate the dipole moment \( \mu \) using the point charge approximation

\begin{equation}
\boldsymbol{\mu} = \sum_i q_i \cdot \mathbf{R}_i
\label{eq:dipole_moment}
\end{equation}
% Where \( q_i \) is the partial atomic charge on atom \( i \) and \( {R}_i \) is the position vector of atom \( i \) in Cartesian coordinates. The summation runs over all atoms in the molecule, and the resulting dipole magnitude \( \mu \) is calculated as the weighted sum of charges and positions. For this method we use QM9, a widely used quantum chemistry dataset comprising approximately 134{,}000 small organic molecules containing carbon, hydrogen, oxygen, nitrogen, and fluorine, with up to nine heavy atoms. QM9 provides a diverse set of quantum mechanical properties computed at the B3LYP/6-31G(2df,p) level of DFT theory for molecules at their equilibrium geometries. We also include QMugs as a secondary benchmarking dataset similar to QM9 which provides a diverse set of quantum mechanical properties for molecules at their equilibrium geometries. QMugs offers a large scale collection of approximately 2 million drug like molecules, with molecules containing tens of heavy atoms, up to about 100, with quantum properties computed at the ωB97X-D/def2-SVP level of DTF using conformers pre-optimized with GFN2-xTB.

where \( q_i \) is the partial atomic charge on atom \( i \) and \( R_i \) is the position vector of atom \( i \) in Cartesian coordinates. The summation runs over all atoms in the molecule, and the resulting dipole magnitude \( \mu \) is calculated as the weighted sum of charges and positions. We employ QM9 dataset, a widely used quantum chemistry dataset comprising approximately 134{,}000 small organic molecules containing carbon, hydrogen, oxygen, nitrogen, and fluorine, with up to nine heavy atoms. QM9 provides a diverse set of quantum mechanical properties computed at the B3LYP/6-31G(2df,p) level of DFT theory for molecules at their equilibrium geometries. We also include QMugs as a secondary benchmarking dataset which provides a diverse set of quantum mechanical properties for molecules at their equilibrium geometries. QMugs offers a large scale collection of approximately 2 million drug like molecules, with molecules containing up to ~100 non-hydrogen atoms, with quantum properties computed at the \(\omega\)B97X-D/def2-SVP level of DTF using conformers pre-optimized with GFN2-xTB. Using the point charge approximation in equation~\eqref{eq:dipole_moment} with Mulliken charges on QM9 yields weak accuracy, with MAE \(= 0.114883\), RMSE \(= 0.143154\), and a mean relative error of \(30.75\%\). This highlights the limitations of Mulliken charges, which stem from the simplified assumptions underlying their computation relative to other charge partitioning schemes. Though they can still be used to help a model learn a more physical, local understanding of charge distributions relevant for dipole prediction. There are other forms of charge assignment schemes that recover the molecular dipole in equation~\eqref{eq:dipole_moment} and are significantly more accurate. In particular, the MSK electrostatic-potential–fit model (Merz–Singh–Kollman)~\cite{SinghKollman1984,BeslerMerzKollman1990} and the more recent ACA machine learning approach~\cite{Sifain2018,Nebgen2018} yield atomic charges that reproduce the molecular dipole. In this work, we use an ML model that incorporates both dipole magnitudes and atomic charges in a loss function to guide the model toward learning to a more physically meaningful state through per node localized electronic features that enhance the prediction of global molecular properties. We intentionally choose Mulliken charges to investigate if addition of labels that lack quantitative accuracy but provide qualitative physics as a secondary task in multitask learning can improve the overall physicality and accuracy of a model. Our motivation is to bypass computationally expensive quantum chemistry calculations -- the construction of the one electron density matrix ${P}$, from which dipoles are obtained as an expectation value of the dipole operator and instead learn this relationship from data for greater computational efficiency at inference. During training, the model predicts atomic charges as a label with atomic numbers and positions as features, and then calculates dipole magnitudes as a label with charges and positions as features. This joint learning objective is represented as:

% In this work, we use an ML model that incorporates both dipole magnitudes and atomic charges in a loss function to guide the model toward learning to a more physically meaningful state through per node localized electronic features that enhance the prediction of global molecular properties. During training, the model predicts atomic charges as a label with atomic numbers and positions as features, and predicts dipole magnitudes as a label with charges and positions as features. This joint learning objective is represented as.

\begin{equation}
\mathcal{L}_{\text{dipole}} = \sqrt{ \frac{1}{M} \sum_{i=1}^{M} (\hat{\mu}_i - \mu_i)^2 } + \frac{1}{M} \sum_{i=1}^{M} |\hat{\mu}_i - \mu_i|
\label{eq:dipole_loss}
\end{equation}

\begin{equation}
\mathcal{L}_{\text{charge}} = \sqrt{ \frac{1}{N} \sum_{j=1}^{N} (\hat{q}_j - q_j)^2 } + \frac{1}{N} \sum_{j=1}^{N} |\hat{q}_j - q_j| + C
\label{eq:charge_loss}
\end{equation}

In Equation~\ref{eq:dipole_loss} (Dipole Loss), \( \hat{\mu}_i \) and \( \mu_i \) represent the predicted and true dipole magnitudes for molecule \( i \), with \( M \) denoting the number of molecules in the batch.  
In Equation~\ref{eq:charge_loss} (Charge Loss), \( \hat{q}_j \) and \( q_j \) are the predicted and true partial atomic charges for atom \( j \), where \( N \) is the total number of atoms in the batch. We use both the QM9 and QMugs datasets, which are composed of small and drug-like molecules with neutral equilibrium structures~\cite{ramakrishnan2014quantum, ireland2021qmugs}. A penalty term \( C \) is applied to the loss function to enforce this neutrality:

\begin{equation}
C = \frac{1}{N} \sum_{j=1}^{N} |\hat{q}_j|
\label{eq:charge_penalty}
\end{equation}
 Training to both charge and dipole is performed together using a composite loss function, where the charge loss is assigned a smaller weight. Although its contribution is relatively minor, the charge loss helps guide the model toward learning more physically consistent representations during training, validation, and testing. The total loss is computed as a weighted sum of the individual loss components.

\begin{equation}
\mathcal{L}_{\text{total}} = \lambda_1 \cdot \mathcal{L}_{\text{charge}} + \lambda_2 \cdot \mathcal{L}_{\text{dipole}}
\label{eq:total_loss}
\end{equation}

In this work, we use Hierarchical Interacting Particle Neural Network (HIP-NN), where features are updated through a sequence of hierarchical layers~\cite{lubbers2018hierarchical}. HIP-NN is specifically designed for learning QM properties of molecules by building up representations of atomic spatial environments through interactions at multiple spatial scales. The network begins by initializing atom wise features based on atomic number, and then iteratively refines them through a hierarchy of interaction layers. We use a three layer architecture: (i) the first layer captures element wise baseline contributions; (ii) the second layer encodes atom atom interactions within the cutoff by expanding interatomic distances on Gaussian basis functions; and (iii) the third layer captures second order, neighborhood to neighborhood interactions. All interactions are restricted by the cutoff \(R_{\text{cut}}\), which sets the maximum distance over which atoms exchange information.

\begin{equation}
f_{\text{cut}}(r) =
\begin{cases}
\left[ \cos\left( \frac{\pi}{2} \cdot \frac{r}{R_{\text{cut}}} \right) \right]^2, & \text{if } r \leq R_{\text{cut}} \\
0, & \text{if } r > R_{\text{cut}}
\end{cases}
\label{eq:cutoff_function}
\end{equation}

The cutoff function \( f_{\text{cut}}(r) \) smoothly reduces interactions to zero as the interatomic distance \( r \) approaches \( R_{\text{cut}} \), ensuring that only nearby atoms contribute to the message passing. In practice, atoms that are very close together contribute strongly, while those near the cutoff distance contribute only weakly, and atoms beyond \( R_{\text{cut}} \) make no contribution at all. This smooth decay avoids discontinuities at the cutoff and helps stabilize the learning process by ensuring physically realistic locality of interactions.

% \newpage

\begin{figure}[hptb]
    \centering
    % Panel (a)
    \begin{subfigure}[b]{0.48\textwidth}
        \centering
        \includegraphics[width=\linewidth]{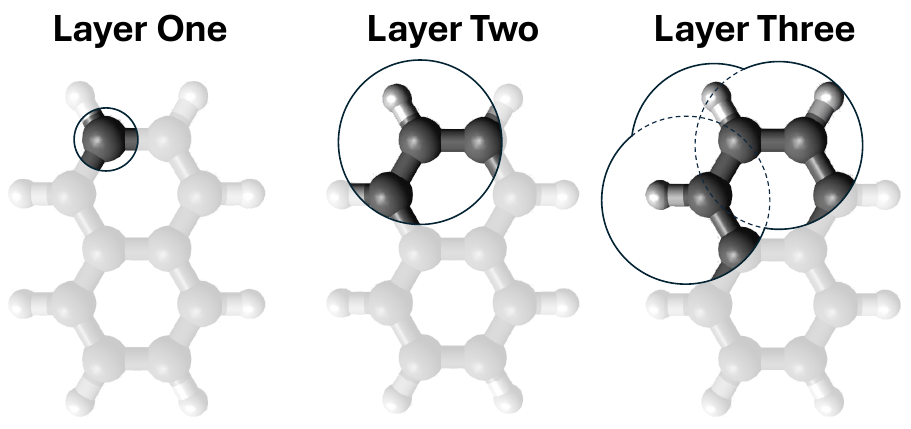}
        \caption{Visualization of cutoff neighborhoods on naphthalene interaction layers communicate within circular regions bounded by \(R_{\mathrm{cut}}\), expanding to larger receptive fields across layers.}

        \label{fig:hip_panel_a}
    \end{subfigure}
    \hfill%
    % Panel (b)
    \begin{subfigure}[b]{0.48\textwidth}
        \centering
        \includegraphics[width=\linewidth]{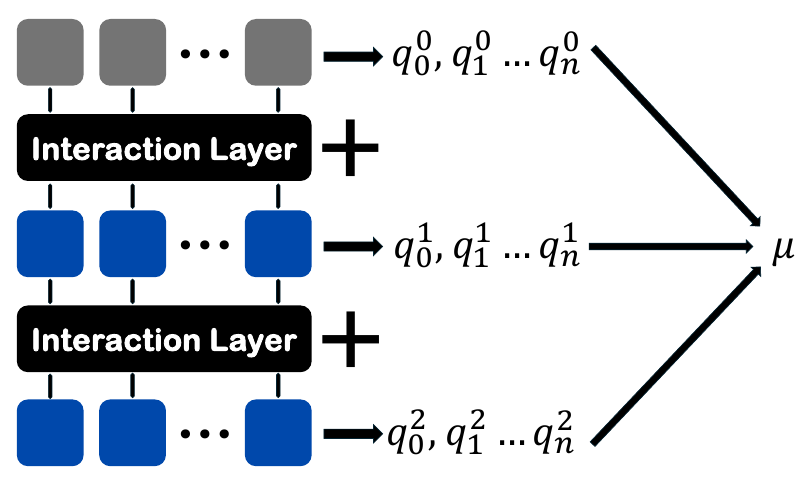}
        \caption{Visualization of interaction layers and there local messages are aggregated and layer outputs are summed into \(\tilde{q}_i\), which is read out to the dipole magnitude \(\mu\).}  
        \label{fig:hip_panel_b}
    \end{subfigure}

    \caption{HIP-NN overview. (a) Layer wise feature extraction. (b) Architecture and aggregation to \(\tilde{q}_i\).}
    \label{fig:hip_combined}
\end{figure}

% \newpage

% \begin{figure}[hptb]
%     \centering
%     \includegraphics[scale=.8]{3.pdf}
%     \caption{
%     Visualization of how hierarchical feature extraction in HIP-NN processes each layer on the small molecule naphthalene.

%     }

%     \label{fig:Davidson_flowchart}
% \end{figure}

% \begin{figure}[hptb]
%     \centering
%     \includegraphics[width=0.65\textwidth]{hip.pdf}
%     \caption{
%         Schematic of the HIP-NN architecture. Each interaction layer outputs atomic features \( q_i^l \) that capture increasingly complex spatial relationships. These outputs are summed across layers to produce the final atom wise representation \( \tilde{q}_i \), which is used to compute the molecular dipole magnitude \( \mu \).
%     }
%     \label{fig:hip_architecture}
% \end{figure}

The contributions from all layers are summed to form the final feature representation \( \tilde{q}_i \), allowing the model to capture both the local chemical environment of each atom and the broader structural context of the molecule. This hierarchical design enables the network to integrate both short range and long range interactions, leading to more accurate predictions of global molecular properties such as dipole magnitude. The contributions from all the layers are summed together to form the final feature representation $\tilde{q}_i$, helping the model understand each atom’s local environment and the overall structure of the molecule as a whole passed through Interaction layers as shown below. This figure is represented in the code as the process that sums the outputs of all interaction layers to form the final atom wise representation \( \tilde{q}_i \). Specifically, the per layer features are aggregated across layers and then passed through a final linear transformation, implemented as a weighted sum with learnable weights and a bias term, as shown in the following equation.

\begin{equation}
\tilde{q}_i = \sum_{a=1}^{N_{\text{features}}} \omega_a^n z_{i,a}^n + b^n
\end{equation}

\noindent
Here, \( \tilde{q}_i \) is the final representation of atom \( i \), \( z_{i,a}^n \) is the \( a^{\text{th}} \) feature of atom \( i \) at layer \( n \), \( \omega_a^n \) are the learnable weights, \( b^n \) is the learnable bias term, and \( N_{\text{features}} \) is the number of feature channels. 
% For this method we use QM9, a widely used quantum chemistry dataset comprising approximately 134{,}000 small organic molecules containing carbon, hydrogen, oxygen, nitrogen, and fluorine, with up to nine heavy atoms. QM9 provides a diverse set of quantum mechanical properties computed at the DFT level of theory for molecules at their equilibrium geometries. We also include QMugs as a secondary benchmarking dataset—similar to QM9 which provides a diverse set of quantum mechanical properties for molecules at their equilibrium geometries. QMugs offers a large scale collection of approximately 2 million drug like molecules, with molecules containing tens of heavy atoms, up to about 100, with quantum properties computed at the DFT level of theory.

\section{Results}

In our benchmarking experiments, we trained and evaluated our machine learning models on two widely used datasets, QM9 and QMugs, using each for both training and testing. We also included a third benchmark to simulate a case where charge values are not available, we do this by computing the average charge for each atom type across the QM9 dataset and use these averages as the charge values. This is denoted as QM9 (Avg) in Table~\ref{tab:charge_table_with_avg}. The assigned charges values are: hydrogen: $+0.1293\,e$, carbon: $-0.0684\,e$, nitrogen: $-0.3014\,e$, oxygen: $-0.3233\,e$, and fluorine: $-0.0657\,e$, where $e$ denotes the elementary charge ($1\,e \approx 1.602 \times 10^{-19}\,\mathrm{C}$). This third baseline allows us to compare model performance in cases where there is approximate, but not exact, electronic environment information for each atom in the molecule. In all models that include charge, we treat it as a secondary learning task where the charge component accounts for only 4\% of the total loss, denoted by $\lambda_1$ in Equation~\eqref{eq:total_loss}.

\subsection{Charge Accuracy}

\begin{figure}[htbp]
    \centering
    \begin{subfigure}[b]{0.48\textwidth}
        \centering
        \includegraphics[width=\textwidth]{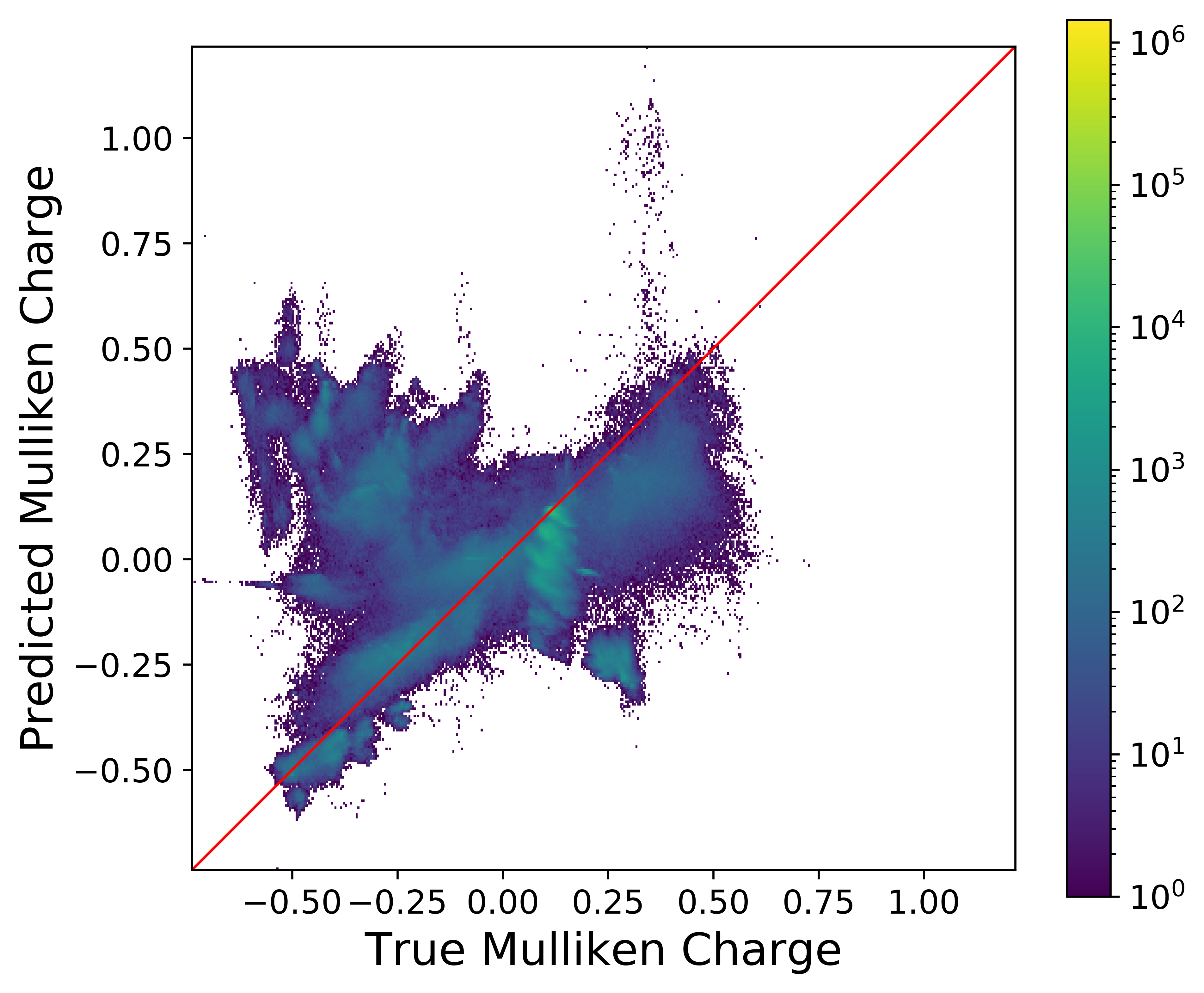}
        \caption{
            Fit when training only to dipoles.
        }
        \label{fig:charge_only}
    \end{subfigure}
    \hfill
    \begin{subfigure}[b]{0.48\textwidth}
        \centering
        \includegraphics[width=\textwidth]{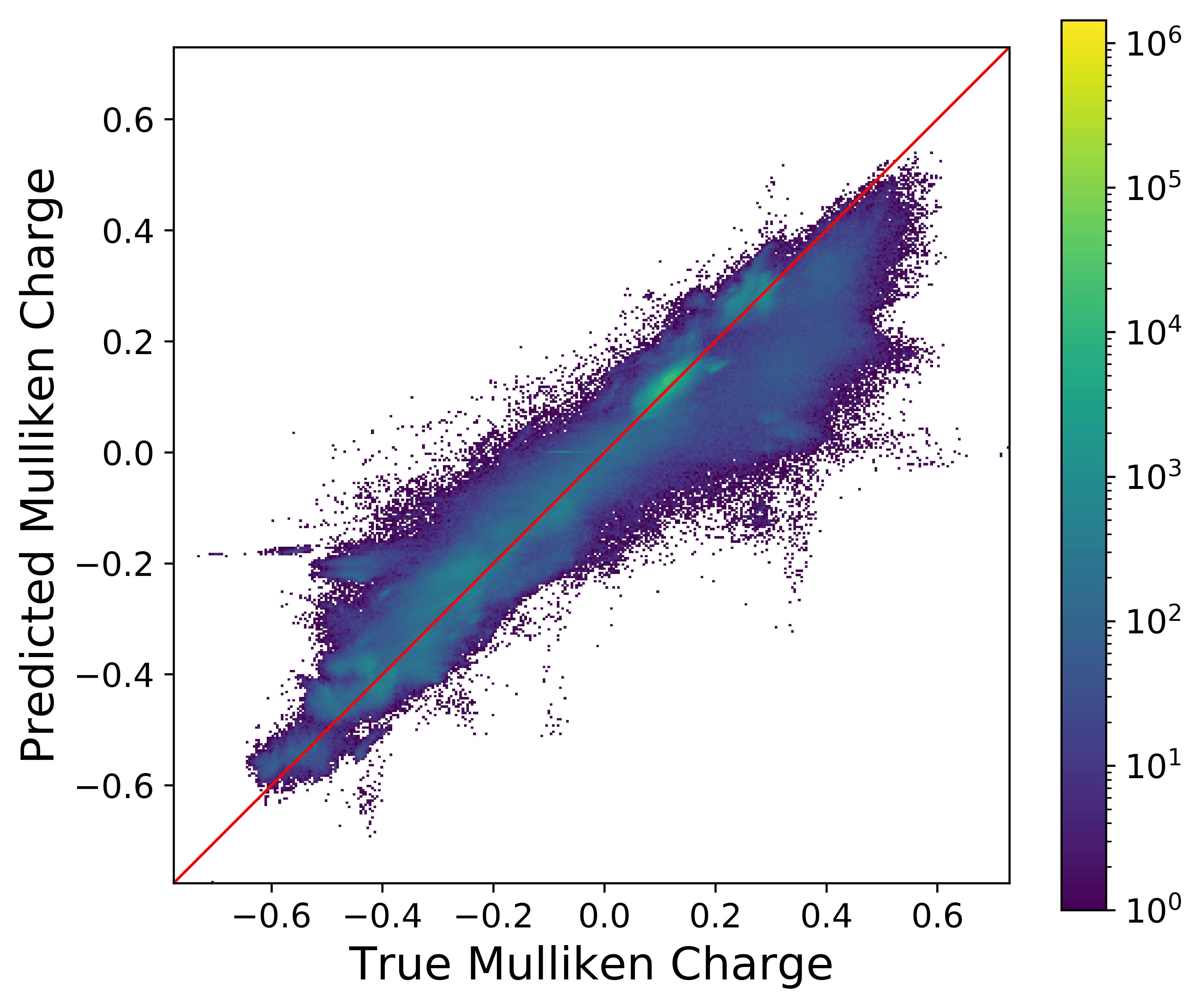}
        \caption{
            Fit when training to dipoles and charges.
        }
        \label{fig:charge_combined}
    \end{subfigure}
    \caption{
                Predicted vs.\ true Mulliken charges on the QM9 dataset. Each point represents a single atom from the validation set, colored by point density on a logarithmic scale. The red diagonal line denotes perfect agreement between prediction and ground truth ($y = x$). 
         }
    \label{fig:charge_comparison2}
\end{figure}

To understand how well the model captures localized electronic information, we evaluate its ability to predict atomic charges under two different training configurations, 1 on a model trained only to fit molecular dipole magnitudes and 2 a multitask training to fit both molecular dipole magnitudes and partial atomic charges. While our primary objective is to predict accurate global dipole magnitudes, this auxiliary task of charge prediction allows us to assess whether the model can learn more physically grounded, atom level electronic features. This is seen in Figure~\ref{fig:charge_comparison2} in the training results and in Table~\ref{tab:charge_table} showing the MAE and RMSE comparisons.

\newpage

Although the RMSE and MAE of predicted charges remain relatively high even when charges are included as a label during training, this is of secondary importance. More importantly, we observe qualitatively that the improved fit indicates the model is beginning to capture more physically meaningful relationships at the atomic scale. The enhanced correlation between predicted and true Mulliken charges suggests that the model is learning a chemically consistent representation of local electronic environments. While not the primary learning objective, the per atom charge predictions act as a useful inductive bias that strengthens the model’s internal representation of molecular charge distributions. This, in turn, equips the model to more reliably infer global molecular properties such as dipole magnitudes, which directly depend on the spatial arrangement of local charges. When atomic charge supervision is excluded, as in \autoref{fig:charge_only}, predictions of true vs. predicted charges vary substantially between runs. The wider spread reflects the fact that the model is not explicitly constrained to reproduce per-atom charges when trained solely on dipole magnitudes. Nevertheless, for both \autoref{fig:charge_only} and \autoref{fig:charge_combined}, a more accurate fit in charge accuracy generally corresponds to improved dipole magnitude accuracy. Interestingly, in \autoref{fig:charge_only}, where charges are not directly supervised, the dipole accuracy improves across many runs when the predicted charges either align with the correct diagonal or fall along the opposite diagonal. The latter case indicates that the model often recovers the correct magnitudes of the charges but assigns the wrong sign. This suggests that, even without explicit charge supervision, the model is still learning chemically meaningful relationships between charges and dipoles, albeit imperfectly.

\begin{table}[h]
\centering
\begin{tabular}{c c c c c}
\toprule
\textbf{Dataset} & & \textbf{Without Charge} & \textbf{With Charge} & \textbf{Improvement} \\
\midrule
\multirow{2}{*}{QM9} & RMSE & 0.254 & 0.061 & 76\% \\
                     & MAE & 0.163 & 0.036 & 78\% \\
\midrule
\multirow{2}{*}{QMugs} & RMSE & 0.6377  & 0.4487  & 30\% \\
                       & MAE  & 0.4358  & 0.3098  & 29\% \\

\bottomrule
\end{tabular}
\caption{Point charge accuracy (compared to Mulliken) on different datasets, where the top value is the root mean squared error and the lower value is the mean absolute error of predicted charges (e).}
\label{tab:charge_table}
\end{table}

\newpage

\subsection{Dipole Accuracy}

\begin{figure}[htbp]
    \centering
    \includegraphics[scale=0.65]{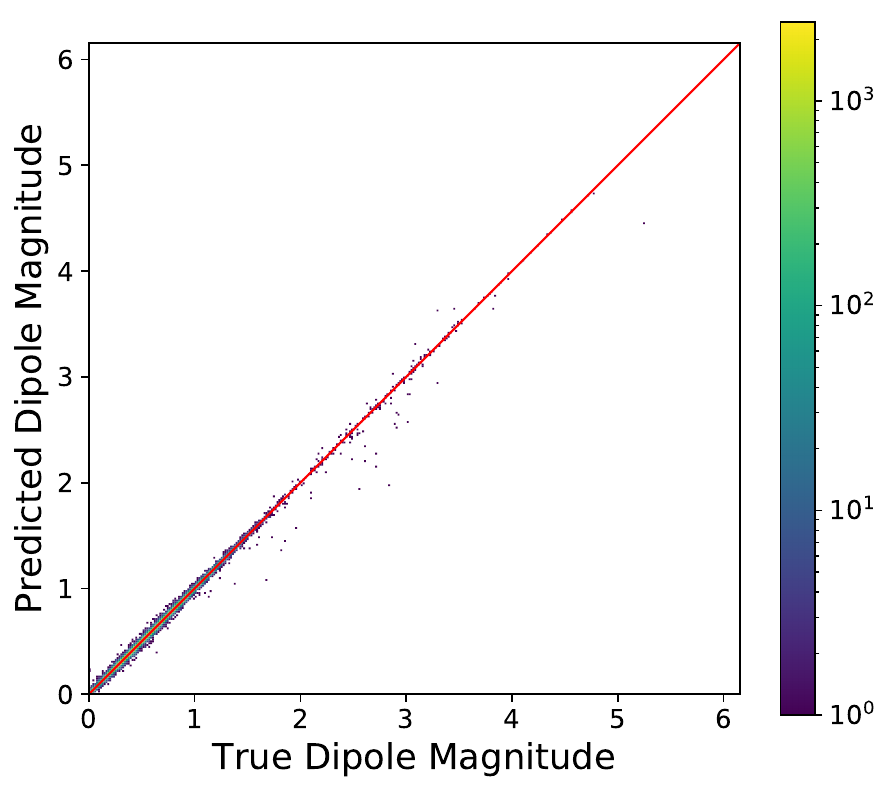}
    \caption{Predicted vs.\ true molecular dipole magnitudes when including charges in the loss function during training on the QM9 dataset. }
    \label{fig:dipole_comparison}
\end{figure}

When training only on dipole magnitudes, we observe a strong fit in the dipole predictions, although predicted point charges are found to be unphysical. However, when atomic charges are included in the loss function, we see a significant improvement in dipole prediction accuracy across benchmarks. This is reflected in Figure~\ref{fig:dipole_comparison}, showing the training results, and in Table~\ref{tab:charge_table_with_avg}, which reports the MAE and RMSE comparisons.

In benchmarking, we observe significant improvements in models trained on both dipoles and charges compared to those trained only on dipoles, as shown in Table~\ref{tab:charge_table_with_avg}. This suggests that the model is learning a more physically grounded representation, resulting in improved accuracy. Across multiple tests with different random seeds, we also observe more consistent performance and fewer extreme outliers when charges are included. Interestingly, we see a improvement even when using average QM9 charge values, indicating that even approximate charge information can help the model learn local and physical interactions more effectively, further enhancing accuracy. It is important to note, however, that Mulliken point charges alone do not yield satisfactory quantitative accuracy for dipole moments. Using the point charge approximation in equation~\eqref{eq:dipole_moment} with Mulliken charges on QM9 results in MAE \(= 0.114883\), RMSE \(= 0.143154\), and a mean relative error of \(30.75\%\), with a systematic overestimation (see SI). This underscores that even when auxiliary data might lack quantitative accuracy, it can provide important qualitative physical insights for the model, improving its performance.

\begin{table}[h]
\centering
\begin{tabular}{c c c c c}
\toprule
\textbf{Dataset} & & \textbf{Without Charge} & \textbf{With Charge} & \textbf{Improvement} \\
\midrule
\multirow{2}{*}{QM9} & RMSE & $0.0246$ & $0.0162$ & 34\% \\
                     & MAE  & $0.0130$ & $0.0098$ & 25\% \\
\midrule
\multirow{2}{*}{QM9 (Avg)} & RMSE & $0.0246$ & $0.01691$ & 31\% \\
                           & MAE  & $0.0130$ & $0.01096$ & 16\% \\
\bottomrule
\end{tabular}
\caption{Comparison of performance and improvement in predicted dipole magnitudes across different datasets, reported in Debye. }
\label{tab:charge_table_with_avg}
\end{table}

% \newpage

We also benchmark against QMugs, a larger and more chemically diverse dataset, and observe improvements when incorporating charge supervision. However, the relative gains are smaller compared to QM9 because the size and diversity of QMugs enable the model trained without charge labels to already capture local electronic environments reasonably well, reducing the added benefit of explicit charge information, as shown in Table~\ref{tab:charge_table_with_avg}. Despite this, the model trained with both dipole and charge information still outperforms the dipole-only model, demonstrating the generalizability and robustness of the multitask learning approach. While the absolute accuracy gains are not as pronounced on QMugs, the multitask framework nevertheless yields a more interpretable and less black box model by leveraging auxiliary charge information to guide the learning process.

\begin{table}[h]
\centering
\begin{tabular}{l c c c c}
\toprule
\textbf{Dataset} & & \textbf{Without Charge} & \textbf{With Charge} & \textbf{Improvement} \\
\midrule
\multirow{2}{*}{QMugs}    & RMSE & $0.3309$ & $0.3103$ & 6\% \\
& MAE  & $0.2131$ & $0.1979$ & 7\% \\
\bottomrule
\end{tabular}
\caption{Comparison of performance on QMugs validation set with and without atomic charges in the loss function, reported in Debye.}
\label{tab:charge_table_with_valid_Qmugs}
\end{table}

\newpage

\section{Conclusion}

In this work, we introduced a multitask ML approach that jointly predicts atomic partial charges and molecular dipole moment magnitudes to improve the accuracy of dipole predictions. Under the assumption that only scalar values of molecular dipole magnitudes are available without their vector components, we investigated the impact of incorporation of lower-quality data -- Mulliken charges -- that on their own do not properly reproduce the target quantum property (dipole moment). Although such data lacks quantitative accuracy, it supplies the model with physical ground. By incorporating a small secondary loss component of atomic charges, we observed a significant improvement in dipole moment magnitude accuracy compared to a model trained only on dipole magnitudes. This improvement stems from the model’s ability to use physically meaningful atomic level information from the charges, allowing it to better capture the effect of local electronic environments and, in turn, predict global extensive molecular properties more accurately. Our results highlight the value of multitask learning in quantum chemistry applications, where auxiliary tasks serve as useful physical features, improving both generalization and interoperability. Notably, we show that even approximate and computationally inexpensive Mulliken charges, despite their inability to properly reproduce quantum dipole moment magnitudes, can still provide significant benefit during training. We only assign a small portion of the loss function to atomic charges as it is important to not focus on fitting to them but instead using them to learn physically meaningful atomic representations. A promising direction for future work could be to extend this multitask framework to incorporate additional quantum chemical properties such as dipole vectors or quadrupole moments, or to apply this approach to non-equilibrium geometries and charged systems, which could open the door to improved simulations under more realistic conditions. It could also be applied to even larger systems, such as large molecules or small proteins with hundreds or thousands of atoms, where long range interactions and collective effects become more important. Overall, our results suggest that multitask learning offers a powerful strategy for improving extensive molecular property prediction by embedding physical insights directly into the learning process. This approach holds promise for advancing data driven molecular modeling and enabling faster, more accurate predictions in computational chemistry, materials science, and drug discovery applications.

\newpage


\begin{acknowledgement}
This work is supported by the U.S. Department of Energy, Office of Basic Energy Sciences (FWP LANLE8AN) and by the U.S. Department of Energy through the Los Alamos National Laboratory (LANL). This work was performed in part at the Center for Integrated Nanotechnology (CINT) at Los Alamos National Laboratory (LANL), a U.S. DOE and Office of Basic Energy Sciences user facility. This research used resources provided by the LANL Institutional Computing Program. LANL is operated by Triad National Security, LLC, for the National Nuclear Security Administration of the U.S. Department of Energy Contract No. 892333218NCA000001. We also thank Emily Shinkle for her helpful contributions to the development of this manuscript.
\end{acknowledgement}

% \begin{suppinfo}

% This will usually read something like: ``Experimental procedures and
% characterization data for all new compounds. The class will
% automatically add a sentence pointing to the information on line:

% \end{suppinfo}

% \bibliography{achemso-demo}
\bibliography{references}
% arXiv won't run BibTeX; include the prebuilt .bbl directly:
\providecommand{\latin}[1]{#1}
\makeatletter
\providecommand{\doi}
  {\begingroup\let\do\@makeother\dospecials
  \catcode`\{=1 \catcode`\}=2 \doi@aux}
\providecommand{\doi@aux}[1]{\endgroup\texttt{#1}}
\makeatother
\providecommand*\mcitethebibliography{\thebibliography}
\csname @ifundefined\endcsname{endmcitethebibliography}
  {\let\endmcitethebibliography\endthebibliography}{}

\end{document}